\documentclass[12pt]{article}
\begin{document}
\def\a{\alpha}
\def\b{\beta}
\def\c{\varepsilon}
\def\d{\delta}
\def\e{\epsilon}
\def\f{\phi}
\def\g{\gamma}
\def\h{\theta}
\def\k{\kappa}
\def\l{\lambda}
\def\m{\mu}
\def\n{\nu}
\def\p{\psi}
\def\q{\partial}
\def\r{\rho}
\def\s{\sigma}
\def\t{\tau}
\def\u{\upsilon}
\def\v{\varphi}
\def\w{\omega}
\def\x{\xi}
\def\y{\eta}
\def\z{\zeta}
\def\D{{\mit \Delta}}
\def\G{\Gamma}
\def\H{\Theta}
\def\L{\Lambda}
\def\F{\Phi}
\def\P{\Psi}
\def\S{\Sigma}

\def\o{\over}
\def\beq{\begin{eqnarray}}
\def\eeq{\end{eqnarray}}
\newcommand{\gsim}{ \mathop{}_{\textstyle \sim}^{\textstyle >} }
\newcommand{\lsim}{ \mathop{}_{\textstyle \sim}^{\textstyle <} }
\newcommand{\vev}[1]{ \left\langle {#1} \right\rangle }
\newcommand{\bra}[1]{ \langle {#1} | }
\newcommand{\ket}[1]{ | {#1} \rangle }
\def\diag{\mathop{\rm diag}\nolimits}
\def\tr{\mathop{\rm tr}}

\baselineskip 0.7cm

\begin{titlepage}

\begin{flushright}
UT-07-30
\end{flushright}

\vskip 1.35cm
\begin{center}
{\large \bf
Supersymmetric Tuned Inflation
}
\vskip 1.2cm
Izawa~K.-I.${}^{1}$ and Y.~Shinbara${}^{2}$
\vskip 0.4cm

${}^1${\it Yukawa Institute for Theoretical Physics, Kyoto University,\\
     Kyoto 606-8502, Japan}

${}^2${\it Department of Physics, University of Tokyo,\\
     Tokyo 113-0033, Japan}

\vskip 1.5cm

\abstract{
We address an issue what kind of tuning
realizes a flat potential for inflaton in supergravity.
We restrict ourselves to the case that inflationary dynamics is not
affected by separate supersymmetry-breaking effects,
and consider examples of a single chiral superfield as an inflaton
with its vacuum of the vanishing cosmological constant.
The examples potentially include the cases of
Polonyi field or right-handed sneutrino as an inflaton.
}
\end{center}
\end{titlepage}

\setcounter{page}{2}

\section{Introduction}

Slow-roll inflation
\cite{Lyt}
takes place when the potential
is flat enough,
which is realized through tuning the form of potential
by symmetry or by hand.%
\footnote{
Fine-tuning problem in inflationary models seems subtle.
Although required parameter tuning itself tends to be unnatural
in field theory landscape,
the tuning is possibly advantageous to realize
(infinitely) large volume of (our) universe,
which may be environmentally favorable and 
compensate unnaturalness of the delicate parameter choices.
For a positive use of such a tuning relevant to particle physics, see 
Ref.\cite{Ibe}.}
In non-supersymmetric or supersymmetry-broken%
\footnote{
See Ref.\cite{All}
for use of separate supersymmetry-breaking effects
(besides inflaton (temporary) supersymmetry breaking) to induce inflation.}
cases, the required tuning is rather straightforward
to flatten a potential with the kinetic term intact.
On the other hand, in the supersymmetry-invariant case,
the K{\" a}hler potential is to be tuned,
which affects not only the potential
but also the kinetic function in supergravity.
This complicates the necessary tuning to induce supergravity inflation.

In this paper, we address an issue what kind of tuning
realizes a flat potential for inflaton in supergravity.
We restrict ourselves to the case of supersymmetry intact.
Namely, we assume that inflationary dynamics is not
affected by separate supersymmetry-breaking effects
if any.

The supersymmetric inflation models with chiral superfields
may be characterized by the number of superfields and the lowest monomial
degree of their superpotential around the vacuum.
We examine examples of a single chiral superfield%
\footnote{See Ref.\cite{Izak} for an example of multi-field case.}
as an inflaton,
for simplicity, with its vacuum of the vanishing cosmological constant.
The examples potentially include the cases of
Polonyi field or right-handed sneutrino as an inflaton.

The rest of the paper goes as follows:
In section 2, we recapitulate the form of scalar potential
in supergravity.
Then we deal with two cases of inflation
with large and small field variations in turn.
In section 3, a tuning for large-field inflation
in effective field theory
is applied to the case of supergravity
and quadratic potentials for chaotic inflation
are obtained.%
\footnote{
See Ref.\cite{Gon,Mur}
for some realizations of chaotic inflation in supergravity.}
The inflaton turns out to be a Polonyi field%
\footnote{
See Ref.\cite{Iza}
for supersymmetry-breaking inflation.}
in one class of models
and a `right-handed sneutrino'%
\footnote{
See Ref.\cite{Mur}
for sneutrino inflation.}
in another, though we do not try to construct
realistic models corresponding to those fields
in this paper.
Section 4 provides
`saddle point'
\cite{All}
models as examples of small-field inflation
with a wider range of physical parameter scales
compared to the above large-field case.
Section 5 concludes the paper.

\section{The supergravity potential}

Let us adopt a chiral superfield $\phi$
with a superpotential $W(\phi)$
and a K{\" a}hler potential
\beq
 K=-3\ln \left( -{\Omega \o 3} \right),
\eeq
where $\Omega = -3 + \cdots$ denotes a real function of $\phi$
and the reduced Planck unit $M_G = 2.4\times 10^{18} {\rm GeV}=1$ is assumed.

The lowest component of the chiral superfield $\phi$
with the abuse of notation consists of two real scalar fields
as $\phi = x + iy$.
Note
\beq
 \Omega_\phi = {1 \over 2}(\Omega_x - i\Omega_y), \quad
 \Omega_{\phi {\phi^*}} = {1 \over 4}(\Omega_{xx} + \Omega_{yy}),
\eeq
where the subscripts denote differentiations with respect to them.
The kinetic function is given by
\beq
 K_{\phi {\phi^*}} = {3 \over \Omega^2}
 (|\Omega_\phi|^2 - \Omega_{\phi {\phi^*}} \Omega)
 \label{kinetic}
\eeq
with $K_{\phi} = -3\Omega_\phi / \Omega$.
The scalar potential in supergravity is given by
\beq
 V &=& e^K(K_{\phi \phi^*}^{-1}|W_\phi+K_\phi W|^2 - 3|W|^2)
 \nonumber \\
 &=&
 -\left( {3 \o \Omega} \right)^3
 \left( {\Omega^2 \o 3(|\Omega_\phi|^2
 - \Omega_{\phi {\phi^*}} \Omega)}
 \left| W_\phi - {3\Omega_\phi \o \Omega}W \right|^2
 - 3|W|^2 \right)
\eeq
for the kinetic function
Eq.(\ref{kinetic}).

\section{Large-field inflation}

A tuning for large-field inflation can be given by
making the kinetic term large
\cite{Dim},
or the overall coupling small.
Let us first consider a Lagrangian
\beq
 {\cal L} = {1 \o 2}A^2\q_\mu {\tilde \v} \q^\mu {\tilde \v} - {\tilde V}({\tilde \v})
\eeq
of a real scalar field $\tilde \v$ with a positive constant $A$.
For the canonical field%
\footnote{
The possible variation of the canonical field $\v$
is consequently larger than
the reduced Planck scale for large $A$
even if that of the original field $\tilde \v$ is restricted
to be within the reduced Planck scale.}
$\v = A {\tilde \v}$,
\beq
 {\cal L} = {1 \o 2}\q_\mu \v \q^\mu \v - V(\v),
\eeq
where $V(\v)={\tilde V}(\v /A)$.
The corresponding slow-roll parameters
\begin{equation}
 \epsilon = \frac{1}{2} \left( \frac{1}{V} \frac{\partial V}{\partial \varphi} \right)^2, \quad 
\eta =  \frac{1}{V} \frac{\partial^2 V}{\partial \varphi^2}
\end{equation}
are given by
\begin{equation}
 \epsilon = \frac{1}{2 A^2} \left( \frac{1}{\tilde{V}} \frac{\partial \tilde{V}}{\partial \tilde{\varphi}} \right)^2, \quad 
\eta =  \frac{1}{A^2 \tilde{V}} \frac{\partial^2 \tilde{V}}{\partial \varphi^2}.
\end{equation}
The sizes of these parameters are rendered to be small
for a large value of $A$ (and large field value $\v$).

A naive guess to apply the above setup to a supersymmetric case
might be to adopt a K{\" a}hler potential $K = A^2|{\tilde \f}|^2$.
However, this does not work since the potential in supergravity
contains the overall factor $\exp K$, in particular,
which does not flatten out for large $A$.

A resolution we propose is, instead, to adopt a distorted
K{\" a}hler potential like
\beq
 K = 2A^2{\tilde x}^2 + {\tilde y}^2,
\eeq
for example, for a complex field ${\tilde \f}={\tilde x}+i{\tilde y}$.
Then the corresponding potential can be flat along the $y$ direction
for a large value of $A$,
since $K = 2x^2 + (y/A)^2$
for an approximately
canonical field $\f = A{\tilde \f} = x + iy$
with $K_{\f {\f^*}}=1+1/(2A^2)$.
We note that a pre-inflation
\cite{Izaw}
would not drive the field $y$
to its origin since the mass of $y$ is suppressed
by $A^{-1}$ in this setup, which might be advantageous for
the initial condition of $y$ inflation to be realized.

The above tuning method is utilized only as a guiding scheme
in this paper for more general tuning of K{\" a}hler potentials.
Let us classify the superpotential according to
the exponent of the inflaton $\phi$ around the vacuum
$\langle \phi \rangle = 0$ (modulo K{\" a}hler transformations).%
\footnote{
The choice of the origin $\phi = 0$ as the vacuum is just a convention.} 
Note that the K{\" a}hler transformation to change the exponent
is singular and inappropriate for the present purposes.

\subsection{The case with $W=c$}

We omit to present the K{\" a}hler potential in terms of
the original field variable $\tilde \phi$ but
provide it in terms of the rescaled one $\phi$ from the start.
Namely, making the coefficients of the $y$-dependent terms tiny as above,
let us consider, for instance,
\beq
 \Omega = -3 + 2\sqrt{3} x - ax^4 - by^n + \sqrt{3} x^3 n(n-1)by^{n-2}f(y),
\label{eq:model1}
\eeq
where $a$ and $b$ are positive constants and $n (\geq 4)$ is even.
The kinetic function
Eq.(\ref{kinetic})
is nearly one for small $|x|$ and $b$. 
The absence of the quadratic and cubic terms%
\footnote{The absence of the $y$-linear term is a convention.}
guarantees the vanishing cosmological constant in the vacuum
$\langle \phi \rangle = 0$
\cite{Heb},
while the supersymmetry is inevitably broken
so that the inflaton is a Polonyi field
with the gravitino mass $m_{3/2} \simeq |c|$
in this case.

We can take a sufficiently large $a$ and
\beq
 f(y)={2 \o 9+3by^n}-{n(n-1)by^{n-2} \o 12+(nby^{n-1})^2}
\eeq
to stabilize the trajectory $x=0$ with the latter choice
eliminating the $y$-dependent tadpole of $x$.
This is not a genuine tuning to induce inflation but
it simplifies the analysis considerably.
For $x=0$,
\beq
 \Omega = -3 - by^n, \quad
 \Omega_\phi = \sqrt{3} + {i \over 2}nby^{n-1},
 \quad \Omega_{\phi {\phi^*}} = -{1 \over 4}n(n-1)by^{n-2},
\eeq
and due to small $b$, for up to moderately large $|y|$,
\beq
 V = {81 |c|^2 \over \Omega^2}\ \frac{\Omega_{\phi {\phi^\dagger}}}
     {\Omega_{\phi {\phi^\dagger}}\Omega - |\Omega_\phi|^2}
   \simeq {3 \over 4}|c|^2n(n-1)by^{n-2},
\eeq
which turns out to be suitable for chaotic inflation,
as is intended.

As for the case of the primordial inflation, the COBE normalization
\cite{Lyt}
implies
\beq
 \left| {V(\v_{N_0})^{3 \over 2} \over V'(\v_{N_0})} \right|
 = 5.3 \times 10^{-4}
 \label{COBE}
\eeq
for a canonical field $\v = \sqrt{2}y$,
where $\v_{N_0}$ is the inflaton field value at the exit
of the present horizon.
For $n=4$, we obtain a quadratic potential with
$b\v_{N_0}^2|c|^2 = 7.5 \times 10^{-7}$,
which implies large gravitino mass $m_{3/2} \simeq |c| \gsim 10^{15}\mathrm{GeV}$.

\subsection{The case with $W=m\phi^2$}

For instance, let us consider
\beq
 \Omega = -3 + dx + \left( 2-{d^2 \o 6} \right) x^2 -ax^4 + x^3f(y).
\eeq
The kinetic function
Eq.(\ref{kinetic})
is nearly one for small $|x|$.

As is the case for the previous example, we can stabilize
the trajectory $x=0$ with the choice
\beq
 f(y) = \frac{d^3}{54} + \frac{8 d (2- y^2)}{48+3 d^2 y^2}.
\eeq
For $x=0$, we obtain
\beq
 \Omega = -3, \quad
 \Omega_\phi = {d \o 2},
 \quad \Omega_{\phi {\phi^*}}
 = 1- \frac{d^2}{12},
\eeq
and the potential is given by
\beq
 V = |m|^2\left( 4y^2+\left\{ \left( {d \o 2} \right)^2-3 \right\}
   y^4 \right).
\eeq

Under a fine-tuning%
\footnote{
For a fixed initial value of the inflaton field
(determined by $b$-dependent terms), the total $e$-fold
is larger in the case of quadratic potential than
that in the case of quartic one. This is a candidate reason
for such a secondary tuning in addition to the primary one of
large $A$, or small $|b|$, to induce inflation.
See also Ref.\cite{Izak}.
}
$d=2\sqrt{3}$ for $2-d^2/6=(d/2)^2-3=0$, 
this potential is quadratic and
the COBE normalization
Eq.(\ref{COBE})
yields
$\v_{N_0}^4|m|^2 = 5.6 \times 10^{-7}$,
which implies $|m| \sim 10^{13}\mathrm{GeV}$.

\section{Small-field inflation}

Small-field models require rather different tuning
to achieve a flat inflaton potential,
which results in inflation with a wider range
of physical parameter scales compared to the previous large-field case.
In the following,
we try to tune the potential of the inflaton $\v$
so that it has a `saddle point'
at $\v = \v_*$ with $|\v_*| \ll 1$,
where the slow-roll parameters satisfy
$\epsilon({\varphi_*}) = \eta({\varphi_*}) = 0$
\cite{All}.
This is a sufficient condition for slow-roll
inflation to be possible. Generally speaking, it is 
unnecessary fine tuning and merely provides
an example of small-field inflation models that
are realized by certain tunings.
The inflaton potential around the `saddle point'
is given by
\begin{equation}
 V(\v) \simeq V(\v_*) + \frac{1}{3!} V'''(\v_*) (\v-\v_*)^3,
 \label{eq:saddlepotential}
\end{equation}
where we assume  $\varphi_*>0$ and $V'''(\varphi_*)>0$ without loss
of generality.%
\footnote{
The initial condition for such inflation might be realized
through primary inflations
\cite{Lyt,Izawa}.}

The inflationary regime ends when the slow-roll condition ($\epsilon<1 $
and $|\eta|<1$) is violated
and such a point $\v=\v_f$ is given by
\begin{eqnarray}
 \v_f \sim \v_* - \frac{V(\v_*)}{V'''(\v_*)}.
\end{eqnarray}
The $e$-fold number $N_e$ corresponding to the value $\v_{N_e}$
of the inflaton field is given by
\begin{eqnarray}
 N_e \simeq \int^{\v_{N_e}}_{\v_f} \frac{V(\v)}{V'(\v)} d \v
 \simeq \frac{2V(\v_0)}{V'''(\v_0)}\frac{1}{\v_*-\v_{N_e}},
\end{eqnarray}
which leads to
\begin{eqnarray}
 \v_{N_e} \simeq \v_* - \frac{2V(\phi_0)}{N_e V'''(\phi_0)}.
 \label{horixon}
\end{eqnarray}

The amplitude of density fluctuations is given by
\begin{eqnarray}
 \left| \frac{V(\v_{N_0})^{\frac{3}{2}}}{V'(\v_{N_0})} \right| 
 \simeq \frac{2V(\v_0)^{\frac{3}{2}}}{V'''(\v_{N_0})(\v_{N_0}-\v_*)^2},
 \label{amplitude}
\end{eqnarray}
where ${N_0}$ denotes the $e$-fold of the present horizon.
The COBE normalization Eq.(\ref{COBE}) thus implies
\begin{eqnarray}
 {V'''(\v_*)^2 \o V(\v_*)} \simeq 1.1 \times 10^{-6} N_0^{-4}.
 \label{cobes}
\end{eqnarray}

\subsection{The case with $W=c$}

Let us consider, for an illustration,
\begin{eqnarray}
 \Omega = -3+ 2\sqrt{3}x - ay^4 + by^5 - 2dy^6,
\end{eqnarray}
where $a$, $b$, $d$ are positive constants.
The kinetic function
Eq.(\ref{kinetic})
is nearly one for small $|x|$ and $|y|$.
The absence of the quadratic and cubic terms implies
that the origin $\phi = 0$ yields a supersymmetry-breaking
minimum with the vanishing cosmological constant
\cite{Heb},
as does in the previous section.

For $x=0$,%
\footnote{As is the case in the previous section,
additional higher-dimensional terms in $\Omega$ are required to
stabilize this trajectory, which we do not show explicitly
any more in this section.}
the scalar potential is given by

\beq
 V = {81|c|^2 \over \Omega^2}\ \frac{\Omega_{\phi {\phi^\dagger}}}
     {\Omega_{\phi {\phi^\dagger}}\Omega - |\Omega_\phi|^2}
   \simeq 3|c|^2{3ay^2-5by^3+15dy^4 \o 1-(3ay^2-5by^3+15dy^4)}.
\eeq
For $b \ll d$, 
this potential possesses a `saddle point' at
\begin{equation}
 y_*={b \o 8d} \ll 1
 \label{phizero}
\end{equation}
when the following condition is satisfied:
\begin{eqnarray}
 a={5 b^2 \o 32d}.
\end{eqnarray}

Then the potential around the `saddle point' $\v_*=\sqrt{2}y_*$ is given by
\begin{eqnarray}
 V
 \simeq \frac{15b}{16\sqrt{2}} |c|^2 \v^3_*
 + {15b \o 4\sqrt{2}} |c|^2 (\varphi-\varphi_*)^3.
\end{eqnarray}
By means of Eq.(\ref{cobes}), the density fluctuations determine
the overall scale as
\begin{eqnarray}
 |c|^2 \simeq 2.9 \times 10^{-9} N_0^{-4} b^{-1} \v_*^3.
\end{eqnarray}
Thus the gravitino mass is given by
\begin{eqnarray}
 m_{3/2} \simeq |c| \simeq 5.4 \times 10^{-5} N_0^{-2}
 \sqrt{b^{-1} \v_*^3} \sim 5\times 10^{11}\mathrm{GeV} \sqrt{b^{-1} \v_*^3},
\end{eqnarray}
which can be small for small-field inflation.
For example,  $\v_* \sim 10^{-3}$ and 
$m_{3/2} \simeq |c| \sim10\mathrm{TeV}$ are realized by tuning $b \sim
10^6$ and $d \sim 10^8$, which is consistent
to the framework of low-energy effective theory
with a cutoff scale larger than $\v_*$.

Finally, we comment on cosmological problems. In this model, the
inflaton mass is given by $m_\varphi \simeq 3 \sqrt{a} m_{3/2}$.
The parameter choice $a \lsim 1$
causes the Polonyi problem for the gravitino mass $\lsim$ 10TeV
when the inflaton decay is Planck suppressed.
For $a > 4/9$, gravitinos
tend to be overproduced through the inflaton decay
even if the gavitino is as heavy as of order 100TeV.
Thus we are led to consider superheavy or superlight gravitino
unless entropy is produced
to dilute the primordial gravitino or
the inflaton decay is enhanced, which may be also advantageous
for baryogenesis at higher reheating temperature.

\subsection{The case with $W=m\phi^2$}

Here we consider, for an illustration,%
\footnote{
This respects a parity $\phi \rightarrow -\phi$,
though such a symmetry is unnecessary just as an illustrative model
for small-field inflation.}
\begin{eqnarray}
 \Omega = -3 + 2xyf + 2x^2\left(1 - {1 \o 3}y^2f^2\right);
 \quad f = a - by^2 + dy^4,
\end{eqnarray}
where $a$, $b$, $d$ are positive constants.
The kinetic function
Eq.(\ref{kinetic})
is nearly one for small $|x|$ and $|y|$.

For $x=0$, the scalar potential is given by
\beq
 V = |m|^2\{ 4y^2 - 3y^4 + (y^3 f)^2 \}.
\eeq
For large $b$ with $b \ll d$,
this potential possesses an approximate `saddle point' at
\begin{equation}
 y_*^2={5b \o 14d} \ll 1
 \label{phizero}
\end{equation}
when the following condition is satisfied:
\begin{eqnarray}
 a={25b^2 \o 84d}.
\end{eqnarray}

Then the potential around the `saddle point' $\v_*=\sqrt{2}y_*$ is given by
\begin{eqnarray}
 V
 \simeq \frac{b^2}{6} |m|^2 \v_*^{10} + {5b^2 \o 126}
 |m|^2 \v_*^7 (\varphi-\varphi_*)^3.
\end{eqnarray}
By means of Eq.(\ref{cobes}), the density fluctuations determine
the overall scale as
\begin{eqnarray}
 |m|^2 \simeq 3.2 \times 10^{-6} N_0^{-4} b^{-2} \v_*^{-4},
\end{eqnarray}
that is,
\beq
 |m| \simeq 1.8 \times 10^{-3} N_0^{-2} b^{-1} \v_*^{-2}
 \sim 10^{12}\mathrm{GeV} b^{-1} \v_*^{-2}.
\eeq
For example, $b \sim 10^7$ and $d \sim 10^9$
result in $a \sim 10^5$, $\v_* \sim 10^{-1}$,
and $|m| \sim 10^7\mathrm{GeV}$.

\section{Conclusion}

If supersymmetry and inflation are indeed operative
in nature, supersymmetric inflation is expected to 
be a key ingredient to realize our present universe. 
Then knowledge of the variety of supersymmetric inflation
models serves as a basis for characterizing the
unique model among them chosen by nature.

We have categorized supersymmetric inflation models
of chiral superfields according to the number of
field multiplets and the exponent of the fields
around the vacuum in the superpotential.
In both the cases of large and small field inflations,
we have also provided concrete examples of tuning schemes
to realize flat potentials for slow-roll inflation
in the case of a single chiral superfield.
That is, tuned K{\" a}hler potentials are given,
which result in flat inflaton potentials
in each superpotential so categorized.

The examples of the tuning schemes we consider
are rather different in two cases of large and small
field inflations, where the variation of inflaton
field is large and small, respectively,
during inflation.
In the large-field case, distorted K{\" a}hler
potentials realize chaotic inflation
by means of the monomial superpotentials
with the exponents zero and two. 
In the small-field case, `saddle point' inflation
is realized by tuning K{\" a}hler potentials
under the monomial superpotentials also with the exponents
zero and two.

Although these tuning methods just provide
mere examples of flattening the inflaton potentials,
they at least show that such tunings are indeed possible.
In view of the subtlety of fine-tuning problem
in inflation mentioned in the first footnote,
it might be expected that even these examples 
might have possible positions to describe nature.

Of course, further consideration is necessary
in order to go beyond these somewhat artificial
setups and arrive at realistic particle-physics models of inflaton.
In particular, extensions to multi-field cases (with supersymmetry breaking)
should also be investigated to obtain global view
on the variety of supersymmetric inflation models.

\section*{Acknowledgements}

We would like to thank M.~Ibe and T.T.~Yanagida for valuable discussions.
Y.S.~thanks the Japan Society for the Promotion of Science for
financial support.
This work is supported by the Grant-in-Aid for the 21st Century COE
"Center for Diversity and Universality in Physics"
from the Ministry of Education, Culture, Sports, Science and
Technology (MEXT) of Japan.

\end{document}